\documentclass{jfm} 
\usepackage{amsmath}  
\usepackage{natbib} 
\usepackage{graphicx} 
\usepackage{placeins} 
\ifCUPmtlplainloaded \else 
  \checkfont{eurm10} 
  \iffontfound 
    \IfFileExists{upmath.sty} 
      {\typeout{^^JFound AMS Euler Roman fonts on the system, 
                   using the 'upmath' package.^^J}% 
       \usepackage{upmath}} 
      {\typeout{^^JFound AMS Euler Roman fonts on the system, but you 
                   dont seem to have the}% 
       \typeout{'upmath' package installed. JFM.cls can take advantage 
                 of these fonts,^^Jif you use 'upmath' package.^^J}% 
      } 
  \else 
  \fi 
\fi 
 
\ifCUPmtlplainloaded \else 
  \checkfont{msam10} 
  \iffontfound 
    \IfFileExists{amssymb.sty} 
      {\typeout{^^JFound AMS Symbol fonts on the system, using the 
                'amssymb' package.^^J}% 
       \usepackage{amssymb}% 

      }{} 
  \fi 
\fi 
 
\ifCUPmtlplainloaded \else 
  \IfFileExists{amsbsy.sty} 
    {\typeout{^^JFound the 'amsbsy' package on the system, using it.^^J}% 
     \usepackage{amsbsy}} 
    {\providecommand\boldsymbol[1]{\mbox{\boldmath $##1$}}} 
\fi

\providecommand\bnabla{\boldsymbol{\nabla}}

\newsavebox{\astrutbox} 
\sbox{\astrutbox}{\rule[-5pt]{0pt}{20pt}}

\newcommand{\bc}{\ensuremath{\mathbf{c}}} 
 
\newcommand{\bu}{\ensuremath{\mathbf{u}}} 
\newcommand{\bx}{\ensuremath{\mathbf{x}}}

\newcommand{\ff}{{\mbox{$\mathbf{ f}$}}} 
\newcommand{\MM}{\ensuremath{\mathbf{M}}} 
\newcommand{\bDelta}{\ensuremath{\pmb{\Delta}}} 

\title{Entropic Lattice Boltzmann Method for Large Scale Turbulence Simulation 
} 
 
\author[Iliya V.\ Karlin et al]% 
{I. \ls V. \ls KARLIN$^1$ 
 \thanks{Corresponding author. E-mail: 
    ikarlin@mat.ethz.ch }  \ns 
S. \ls ANSUMALI $^1$ 
E. \ls DE ANGELIS $^1$ 
\thanks{permanent address: Dip. di Meccanica e Aeronautica, Universit\`a di Roma ``La 
  Sapienza'', Via Eudossiana, 18 00184 Roma, Italy} \break 
H. \ls C. \ls \"OTTINGER$^1$ 
\break 
\and \break 
  S. \ls SUCCI $^2$} 
 
\affiliation{$^1$ 
ETH-Z\"urich, Department of Materials, Institute of Polymers, 
ETH-Zentrum, Sonneggstr. 3, ML J 19, CH-8092 Z\"urich, Switzerland\\[\affilskip] 
$^2$Istituto Applicazioni Calcolo, CNR, viale Policnico 137, 00161 
Roma, Italy 
} 
 
\pubyear{1996} 
\volume{538} 
\pagerange{119--126} 
\date{$30$th May 2003} 
\begin{document} 
 
\maketitle 
 
\begin{abstract} 
Recently, a minimal kinetic model for fluid flow, known as 
entropic lattice Boltzmann method, has been proposed  
for the simulation of isothermal hydrodynamic flows.
At variance with previous Lattice Boltzmann methods, the
entropic version permits to describe the non-linear
dynamics of short scales in a controlled and stable way.
In this paper, we provide the first numerical evidence that
the entropic lattice Boltzmann scheme provides a quantitatively
correct description of the large-scale structures of 
two-dimensional turbulence, free of numerical instabilities.
This indicates that the entropic lattice Boltzmann method 
might provide a starting basis for the formulation of a new class
of turbulence models based on genuinely kinetic principles. 
\end{abstract} 
 
\section{Introduction} 
  \{\}
The dynamics of fully developed turbulent flows 
is characterized by the simultaneous non-linear interaction of a large 
number of degrees of freedom, far too many for analytical
treatment and also for the processing 
capabilities of even most powerful computers. 
Since most turbulent flows of practical relevance 
cannot be simulated {\it ab initio}, they must be modeled. 
The goal of such a modeling is to predict the effects of 
unresolved scales of motion, those that cannot be represented
within the computer simulation,  on the resolved ones. 
Mathematically, these effects are represented by the divergence   
of the Reynolds stress tensor, 
$\boldsymbol{\Gamma} = \left< \mathbf{u}' \mathbf{u}'\right> $, 
where $ \mathbf{u}'$ denotes the short-scale component of fluid  
motion, and the pointed  bracket stands for ensemble averaging. 

A popular class of models is based on the notion of `eddy-viscosity', 
according to which the effects of short scales can be 
incorporated within an effective viscosity  
$\nu_e=\nu+\nu_t$ acting upon the large scales. 
Here $\nu$ is the bare, `molecular' viscosity, whereas $\nu_t$ 
denotes the turbulent viscosity. 
The effective viscosity $\nu_e$ is then expressed as a function(al) 
of resolved fields. The simplest option consists of choosing 
a suitable algebraic relation $\nu_t=\nu_t(S)$, where $S$ is the 
magnitude of the shear tensor $\mathbf{S} = \boldsymbol{\nabla}
\mathbf{U}+\boldsymbol{\nabla} (\mathbf{U})^T
$, $\mathbf{U}$ denoting the resolved flow field. 
The simplest eddy-viscosity models express the Reynolds stress 
tensor in an algebraic form: 
%\begin{equation} 
%\label{EDDY} 
$\boldsymbol{\Gamma}  = \nu_e (\mathbf{S}) \mathbf{S}.$
%\end{equation} 
This formulation has the significant advantage of leaving the form of the  
Navier-Stokes equation for the large eddies intact, with  
an effective viscosity $\nu_e$ replacing the molecular viscosity $\nu$.  
However, such an algebraic representation often fails to reproduce
strongly off-equilibrium turbulent effects, such as those occurring
in the vicinity of solid walls.
A better account of these phenomena is obtained by
linking $\nu_e$ to the actual content 
of turbulent kinetic energy $ k = <{\mathbf{u}'}^2/2> $
and energy dissipation rate $\epsilon = dk/dt$, via $\nu_e \sim k^2/\epsilon$ .
Even so, important phenomena such as turbulent
gusts triggered by local instabilities (corresponding to negative effective
viscosity) cannot be captured \cite[]{EddyTwoD} because $\nu_e$
is bound to be positive definite. Modern formulations which allow for negative viscosities and
corresponding backscatter effects in $k$-space have indeed been
developed in the last two decades \cite[]{LM}. However,  to the best of our knowledge,
none of these models has gained universal consensus to date.

All eddy-viscosity models rely on the analogy between the turbulent
transport and the molecular transport. 
Within this analogy, the small-scale eddies play the 
role of the molecules, while the smallest
coherence length, $l_k$, known as Kolmogorov length,
plays the role of the mean-free path. It is then natural to define a `turbulent Knudsen number' for an
eddy of size $l$ as
\[
Kn_t(l) = \frac{l_k}{l}
\]
It is then understood that small-scale eddies of size close to $l_k$
are in local equilibrium with eddies of size $l$, so
long as these two scales are well separated, i.e.  $Kn_t(l)<<1$. 

A major problem with this analogy is that
turbulence features a {\it continuum} spectrum 
of excitations, so that the above scale-separation argument 
simply does not hold.
In particular, from the above definition of turbulent Knudsen number,
it is clear that the local equilibrium assumption becomes
less and less tenable as $l \rightarrow l_k$.
This lack of scale separation is a fundamental problem which lies
the heart of all failures to develop a consistent theory of fully
developed turbulence.

From the computational point of view, the Kolmogorov
length is replaced by the grid spacing $\delta>>l_k$, indicating
that the dynamics of the smallest resolved scales faces
a situation similar to finite-Knudsen flows at $Kn \sim 1$.

It has been recently pointed out that the 
kinetic representation of hydrodynamics provides a 
natural generalization of the notion of eddy-viscosity 
to such non-equilibrium high-$Kn_{\rm t}$ regimes \cite[]{TurbK, SCI}. 
The key point is that solutions of the kinetic equation 
apply at {\it all} orders of the Knudsen number, so that
{\it any} kinetic model ensuring correct hydrodynamic
behaviour would handle the dynamics of small eddies at $Kn_t \sim 1$
in a way which goes beyond the eddy-viscosity representation. 

The practical problem with kinetic theory, however, is the
enormous increase of degrees of freedom associated with
the (true) Boltzmann equation. 
Thus, in order to apply the aforementioned kinetic
approach to fluid turbulence, drastically simplified 
versions of the Boltzmann equation need to be developed. 
The lattice Boltzmann method (hereafter LBM) \cite[]{succi,BSV}, is 
a good candidate for such a task.
Indeed, the method has been  applied to a large variety
of fluid flows, including turbulent ones \cite[]{TurbFLBM, TurbLBM}.
However, the  standard lattice Boltzmann method meets with
severe difficulties in handling the dynamics of near-grid
scales with $l \sim \delta$. In particular, owing to the lack of
a H-theorem, these scales often exhibit disruptive non-linear
instabilities.

Recently, a new approach to stabilize the kinetic scheme based 
on thermodynamic considerations was developed \cite[]{DHT,
AK2, AK3, AK5, BOG}.    
The modified method known as ``entropic Lattice Boltzmann method''  might
add a further boost towards the formulation of a kinetic
model of fluid turbulence. In particular, its non-linear stability,
and the ensuing positivity of the distribution function, automatically
enforces an important realizability constraint which is missing 
in standard Lattice Boltzmann representations.  
More specifically, this model provides a local, adaptive, regulator of
the relaxation time, which can be regarded to all effects and purposes
as a turbulence model inspired by genuinely kinetic requirements.
In this paper, we present a set of numerical experiments 
which are intended to test this conjecture in a 
semi-quantitative sense. 

The work is organized as follows: first, a  description of the 
LBM and its entropic version is introduced Sec. \ref{Sec: LBE}). 
Then, simulations of two dimensional decaying turbulence
are presented for the fully resolved as well as unresolved
case (Sec. \ref{Sec: DT}). For validation purposes,
results are compared with spectral simulation of the
Navier--Stokes equations in the fully resolved case. 

%In the unresolved case, for the first time an LBM simulation, without 
%any explicit turbulence modeling, of very high Reynolds
%number flow will be presented. 
%This indicates that the entropic lattice Boltzmann method 
%acts like an built-in turbulent model. 
 
\section{Lattice kinetic theory \label{Sec: LBE}} 
 
In the last decade, it has become apparent that minimal versions 
of the Boltzmann kinetic equation provide a fairly efficient alternative 
to the discretisation of the Navier-Stokes equations for the simulation 
of a variety of complex flows, including turbulent ones \cite[]{succi}. 
In its simplest instance, this minimal kinetic equation takes the 
following form: 
\begin{equation} 
\label{LBM} 
\partial_t f_i+ \bc_{i} \cdot \partial_{\bx}f_i = 
-\tau^{-1} \left(f_i- f_i^{\rm eq}  \right), 
\end{equation} 
where 
$f_i \equiv f(x,\bc_{i},t)$ denotes the probability of finding a 
particle at position $x$ and time $t$, moving along the discrete 
direction $\bc_{i}$. The right-hand-side of this equation 
represents collisional relaxation to a local equilibrium  
on a time scale $\tau$.  
The discrete velocities must be chosen in such a way as to 
ensure sufficient symmetry to recover the basic  
mass, momentum and momentum-flux conservations, so that the 
Navier-Stokes equations are recovered as the large-scale limit of  
the discrete kinetic equation. For this purpose, many options have 
become available over the years. 
The major appeal of the kinetic approach is its remarkable simplicity, 
which translates into a corresponding computational efficiency.  
A distinctive mark of this simplicity is 
the fact that, $i$) the streaming operator is linear and proceeds along 
a set of constant directions, $ii$) the non-linearity is entirely due
to  the local equilibrium, which is fully local in configuration space. 
This is in vivid contrast to the Navier-Stokes representation in which 
both non-locality and non-linearity are condensed into a single term, namely 
the convective operator $\mathbf{u} \cdot \partial_{\bx} \mathbf{u}$. 
Typically, the local equilibrium is chosen in the form of a quadratic
polynomial in the fluid speed : 
\begin{equation} 
\label{LEQ} 
f_i^{eq}=  \rho \,  W_{i}\biggl[
1+ k_1 \bc_i \cdot \bu + k_2 
\left(\bc_i \cdot \bu \right)^2 + k_3  (\bc_i \bc_i):( \bu  \bu) \biggr],
\end{equation} 
where, $W_{i}, k_1, k_2$, and $k_3$ are lattice 
dependent constants depending on the lattice sound speed $c_s$ (a constant
for the present case of athermal flows).  
The advantage of this approach is that local equilibria can be readily 
constructed  after particles have gone through the streaming phase. 
Subsequent irreversible relaxation to this local equilibrium provides 
viscous behavior, with a kinematic viscosity of the order of $\nu \sim c_s^2 \tau$. 
The simplicity of this scheme is hard to beat.  
However, the fact that local equilibria are prescribed at the outset and do not 
result from a self-consistent relaxation dynamics in kinetic space, implies that 
compliance with the second law of thermodynamics, the H-theorem in kinetic 
language, is generally lost \cite[]{RMP}. 
The result is that, whenever large gradients develop on the lattice 
scale, as it is the case for fully developed turbulence, 
standard LBE schemes are subject to numerical instabilities. 
This provides a strong motivation to develop a new class of
lattice Boltzmann schemes, capable of accommodating the $H$ theorem. 
A particularly interesting option has recently emerged 
in the form of the  so-called entropic Lattice Boltzmann method (ELBM) \cite[]{AK5}. 
In remainder of this section we will describe entropic  model.   

\begin{figure} 
\begin{center} 
\includegraphics[height=.225\textheight]{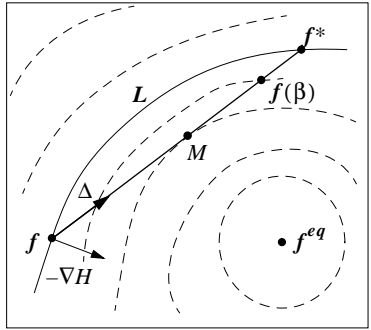} 
\end{center}  
\caption{Stabilization procedure. Curves represent 
entropy levels, surrounding the local equilibrium $\ff_{\rm eq}$. 
The solid curve $L$ is the entropy level with 
the value $H(\ff)=H(\ff^*)$, where $\ff$ is the initial, and $\ff^*$ is 
the auxiliary population. The vector $\bDelta$ represents 
the collision integral, the sharp angle between $\bDelta$ and 
the vector $-\bnabla H$ reflects the entropy production 
inequality. The point $\ff^{*}$ is the solution to Eq. (\ref{Eq:DHT}). The 
result of the collision update is represented by  point 
$\ff(\beta)$.
The choice of $\beta$ shown corresponds to the `over-relaxation': 
$H(\ff(\beta))<H(\ff)$ but $H(\ff(\beta))>H(\MM)$. The particular 
case of the BGK collision (not shown) 
would be represented by 
a vector $\bDelta_{\rm BGK}$, pointing from $\ff$ towards 
$\ff^{\rm eq}$, in which case $\MM=\ff^{\rm eq}$. 
} 
\label{Fig1} 
\end{figure}

The basic strategy of the entropic Lattice Boltzmann 
method is to write the dynamics in terms of a properly 
chosen $H$ function \cite[]{DHT, AK5}. 
For  isothermal hydrodynamics, the discrete $H$ function is found
to be in Boltzmann-form: 
\begin{equation} 
\label{app:H} 
H_{\{W_i,\bc_i\}}=\sum_{i=1}^{b^D} f_{i}\ln\left(\frac{f_{i}}{W_i} \right),
\end{equation} 
where $b^D$ is the number of discrete velocities in $D$-dimensions. The discrete
velocities are tensor products of the discrete velocities in one
dimension  and the weights are constructed by multiplying weights
associated with each  direction.
 The minimal set of discrete velocities needed to reconstruct the
Navier-Stokes equations are related to  zeroes of  third-order
Hermite polynomials in $c_i$ and  
for $D=1$, the three discrete velocities are
$ c_i = c\{-1, 0, 1\}$,  whereas the corresponding weights are
$w = \left \{\frac{1}{6},\frac{2}{3}, \frac{1}{6} \right \}$\cite[]{DLBM}. 
The discrete-velocity local equilibrium is the minimizer of the corresponding 
entropy function under the constraint of local conservation laws:  $\sum_{i=1}^{b^D} f^{\rm eq}_i \{ 1,\  \bc_{i} \}
=\{\rho,\ \rho \bu \}$. The explicit solution for the $f^{\rm
  eq}_i$ is 
\begin{align}
\label{TED}
\begin{split}
f^{\rm eq}_i =\rho W_i\prod_{j=1}^{D} 
%\left[
\left(2 -\sqrt{1+ 3 {u_{j}}^2}\right)
\left(
\frac{ 2 u_{j}+ 
\sqrt{1+ 3 {u_{j}}^2}}{1- u_j}
\right)^{c_{i j}/c},
%\right]
\end{split}
\end{align}
with  $j$ being the index for spatial directions.  
Further, a  notion of the  bare collision $\bDelta$,   
defined  as the collision term stripped of its relaxation parameters
is introduced.  
In the case of BGK model $\bDelta= \ff_{\rm eq}- \ff$, namely 
the bare departure from local equilibrium. 
The time stepping in this method is  performed through an
over-relaxation  collisional process and linear convection  through a sequence 
of steps in which the $H$ function is bound not to decrease.
The  monotonicity constraint on the $H$ function 
is imposed through the following geometric procedure:
In the first step, populations are changed 
in the direction of the bare collision in such a  way that  
the $H$ function remains constant. In the second 
step, dissipation is introduced and the magnitude of the  
$H$ function  decreases. Thus, 
\begin{equation} 
\label{lbe2} 
f_i(\bx,  \delta t)  
=f_i(\bx -\bc_i \delta t,  0)+ 
\alpha \beta  
\biggl[f^{\rm eq}_i (\bx -\bc_i \delta t,  0) - \ff(\bx - \bc_i \delta t,  0) 
\biggr] 
\end{equation} 
were $\beta$ is the discrete form of the relaxation frequency related
to $\tau^{-1}$ and the  parameter $\alpha$ 
is defined by the condition: 
\begin{equation} 
\label{Eq:DHT} 
H\left(\ff\right)=H\left(\ff+{\alpha} \bDelta\right), 
\end{equation} 
and close to the local equilibrium $\alpha$ is equal to $2$ \cite[]{AK2}.  The scheme is illustrated graphically in  
figure \ref{Fig1}. As shown in  figure \ref{Fig1}, in order to find the 
desired point $\ff(\beta)$, we first find the point $\ff^{*}$ on the 
curve $L$ of constant $H$ function.  
This way of enforcing the $H$ theorem  ensures non-linear stability. 
An algorithm to implement  Eq. (\ref{Eq:DHT}) has been
presented by \cite{AK2}.      
The  local adjustments of the
relaxation time (via the parameter $\alpha$), as dictated by compliance
with the H-theorem, guarantee positivity of the distribution function
also for the case of discrete steps, thereby
ensuring non-linear stability of the numerical scheme. 

A comment on the relation between non-linear stability and
built-in subgrid modelling is in order. As is well known, in a
 turbulent flow energy (enstrophy in two-dimensional set up)  cascades from
large to small scales.  
While physically the cascade picture is terminated by dissipation 
at the Kolmogorov length $l_k$, in an underresolved 
simulation, the cascade needs to be
truncated at the grid spacing $\delta>l_k$. As the
entropy of the ELBM (``grid entropy'') accounts for the net effect
of {\it all} degrees of freedom that are not resolved in the simulation
(including the physical entropy), it provides the natural key to
implementing artificial dissipation on the grid scale. According
to the H theorem, the cascade is terminated on the grid scale in
such a way that the subgrid scales cannot ``fire back'' at scales
larger than $\delta$. Consequently, the grid entropy of the ELBM allows us to
decouple all subgrid effects from the cascade in a most natural way.
This  is exactly the goal traditionally approached with
the concept of eddy-viscosity by pushing subgrid effects to the
decoupled level of local equilibrium thermodynamics.

\section{Decaying turbulence \label{Sec: DT}} 
 
In this section, we compare the results of the Entropic 
Lattice Boltzmann simulation with those of pseudo-spectral simulation 
of the Navier-Stokes equations (hereafter SS) for the
case of two-dimensional, homogeneus, incompressible decaying turbulence.
In this setup, one starts with a Gaussian random field in a 
periodic box, and probes the decay of the turbulent field.  

The initial conditions are given by a zero-mean Gaussian vorticity field with 
random Fourier phases.  
The functional form of the  energy spectrum is frequently chosen as: 
%%% 
\begin{equation} 
\label{ICE} 
E(k) = C_0 k^A[1+(k/k_0)^{B+1}]^{-1}, 
\end{equation} 
where, $C_0$ is a normalization constant and   parameters $A$, $B$ and 
$k_0$ are chosen in such a way that the energy spectrum 
remains narrow banded \cite[]{Millen}.  
   
As a SS algorithm, we use the pseudo-spectral method, i.e., the 
non-linear terms are evaluated in 
real space, with a formulation in primitive 
variables (the two velocity components).  
The equation to be solved in Fourier space looks as follows:
\begin{align} 
\label{NSsp} 
\frac{\partial \hat u_i}{\partial t} + j k_j \widehat{(u_j u_i)} = - j k_i \hat{p}  
+ \frac{1}{Re} k^2 \hat u_i 
\end{align} 
where the symbol $\hat{Q}$ denotes the Fourier transform of the 
quantity $Q$.  
At each time step the nonlinear term is projected onto a solenoidal
field in order to  maintain incompressibility condition $k_i \hat u_i = 0$; this procedure is straightforward in $k$-space, being ${\cal P}(j k_j \widehat{(u_j u_i)})=  
j k_j \left(\delta_{il} + {u_i u_l}/{k^2}\right)\widehat{(u_j
  u_l)}$, 
where ${\cal P}$ is the projecting operator. 
The evaluation of the nonlinear terms is performed with a $3/2$-dealiased 
pseudo-spectral method. For the time-marching, a semi-implicit 
low storage third-order accurate Runga-Kutta scheme  
was used \cite[]{splart}. 
 
In the following subsections, a detailed comparison of the result of 
decaying turbulence simulated with SS and ELBM is presented.  
 
\subsection{\label{sect-RS} Fully-resolved simulation} 
 
We consider a two-dimensional box of length   
$L=512$ and viscosity $\nu=5.0 \times 10^{-4}$ (all quantities are
in lattice units), for the SS as well as the ELBM. 
As described earlier, the initial conditions are given in  
Fourier space with $k_0=0.2$, $A=6$ and $B=17$ (see Eq. \ref{ICE}).  
The initial spectrum is reasonably peaked at low wave numbers.  
The initial velocity profile has mean kinetic energy $E(t=0)= 
1.64505947 \times 10^{-4}$ and the mean enstrophy is $Z(t=0)= 
2.0348495\times 10^{-6}$. A rough estimate of the eddy turnover time is $t_e 
\approx Z^{-1/2}  \approx 700$ \cite[]{Millen}. The Reynolds number 
based on the mean initial kinetic energy and the box-length as the 
characteristic length is $Re = L \sqrt{(2 E)} / \nu \approx 
13134$. The estimated dissipation length is 
$L_{d}\sim L \, Re^{-1/2} \approx 3.2$. 
 
\begin{figure} 
\begin{center} 
\includegraphics[scale=0.25]{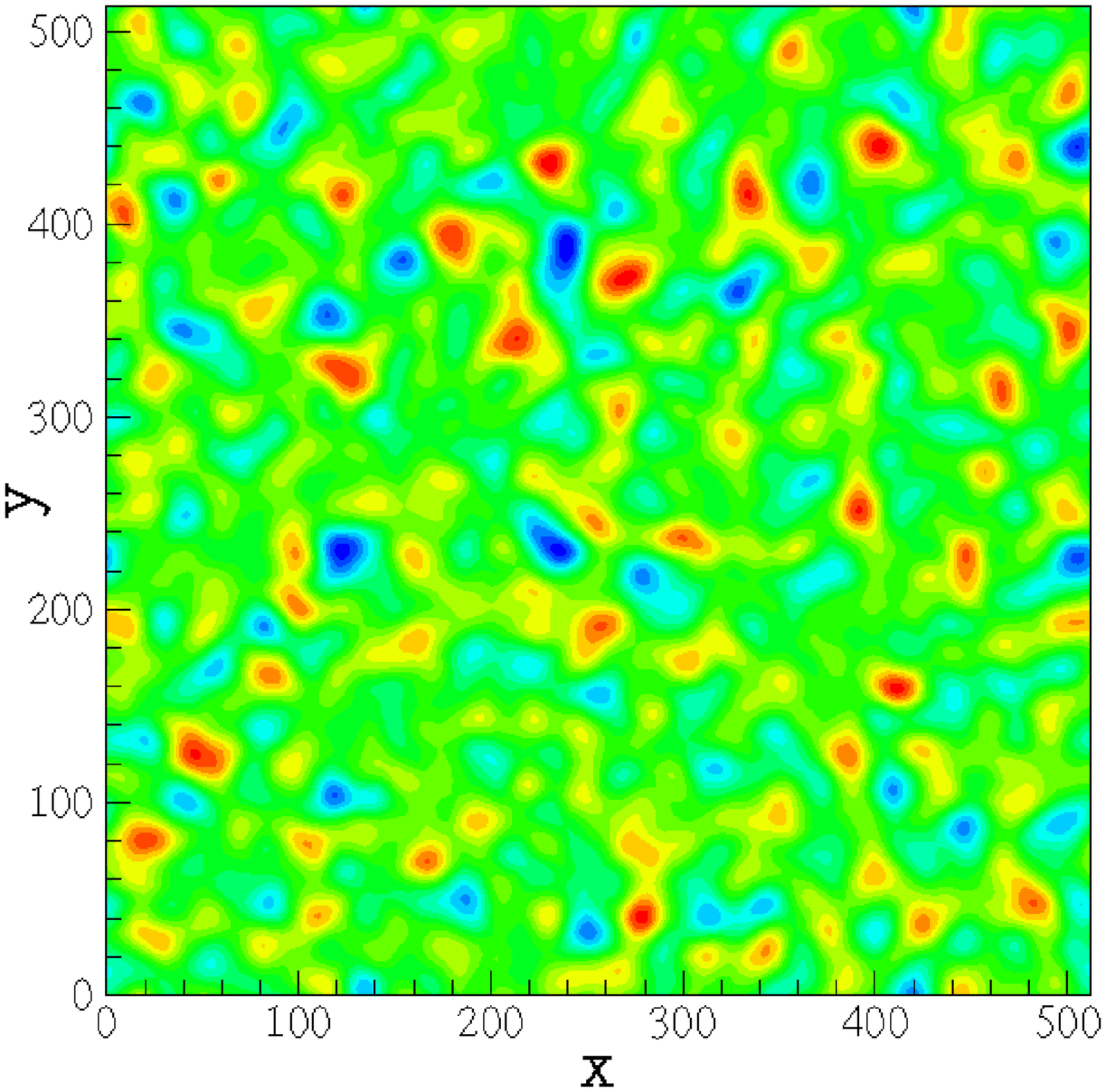} 
\hspace*{0.9cm} 
\includegraphics[scale=0.25]{pics/vortAnsu0000.ps} 
\end{center} 
\vspace*{-1.0cm}  
%\begin{center} 
%\includegraphics[scale=0.2743]{pics/vortElisa1000.ps} 
%\hspace*{0.9cm} 
%\includegraphics[scale=0.3]{pics/vortAnsu1000.ps} 
%\end{center} 
\vspace*{-1.0cm}  
\begin{center} 
\includegraphics[scale= 0.2285833]{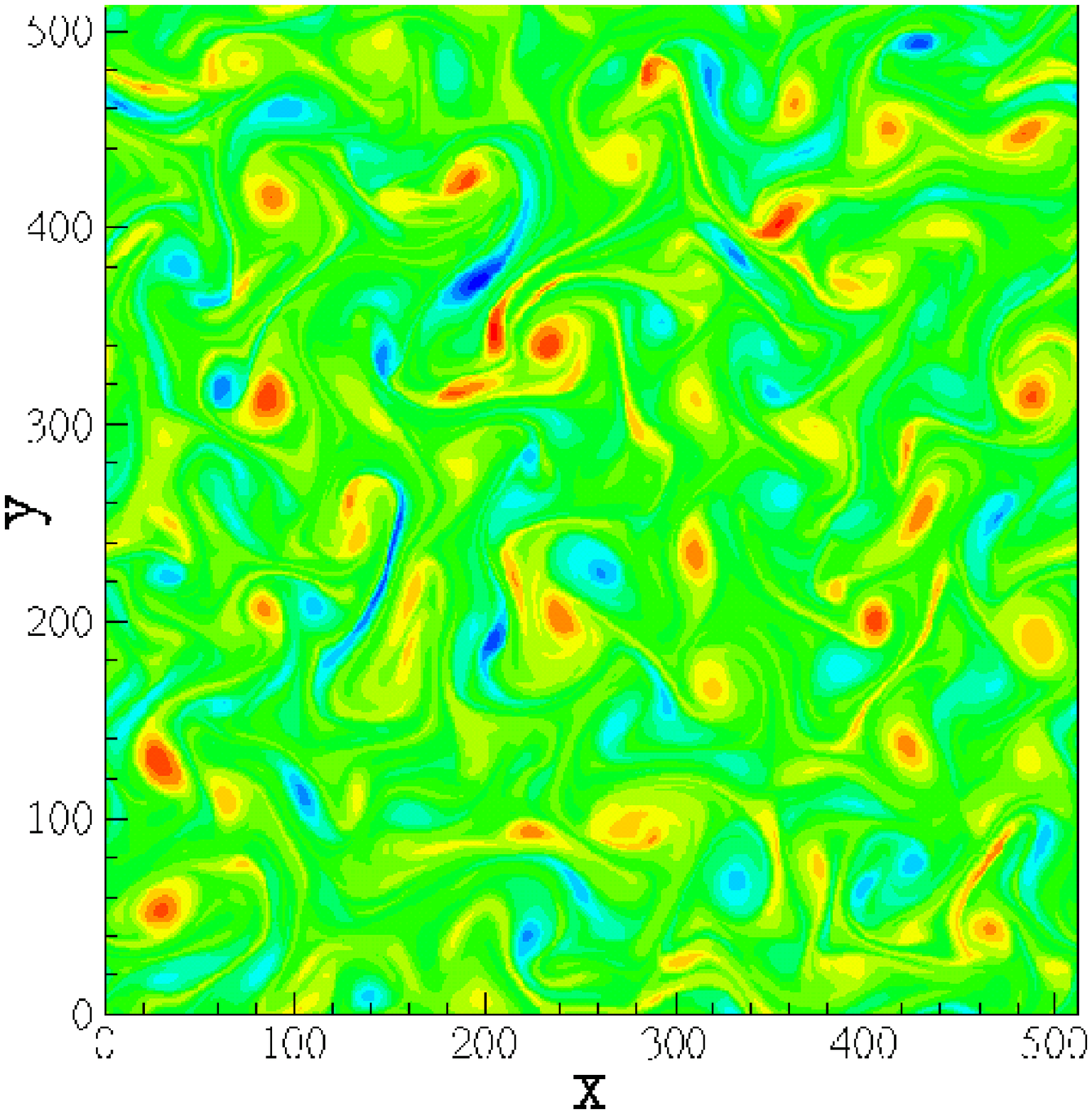} 
\hspace*{0.9cm} 
\includegraphics[scale=0.25]{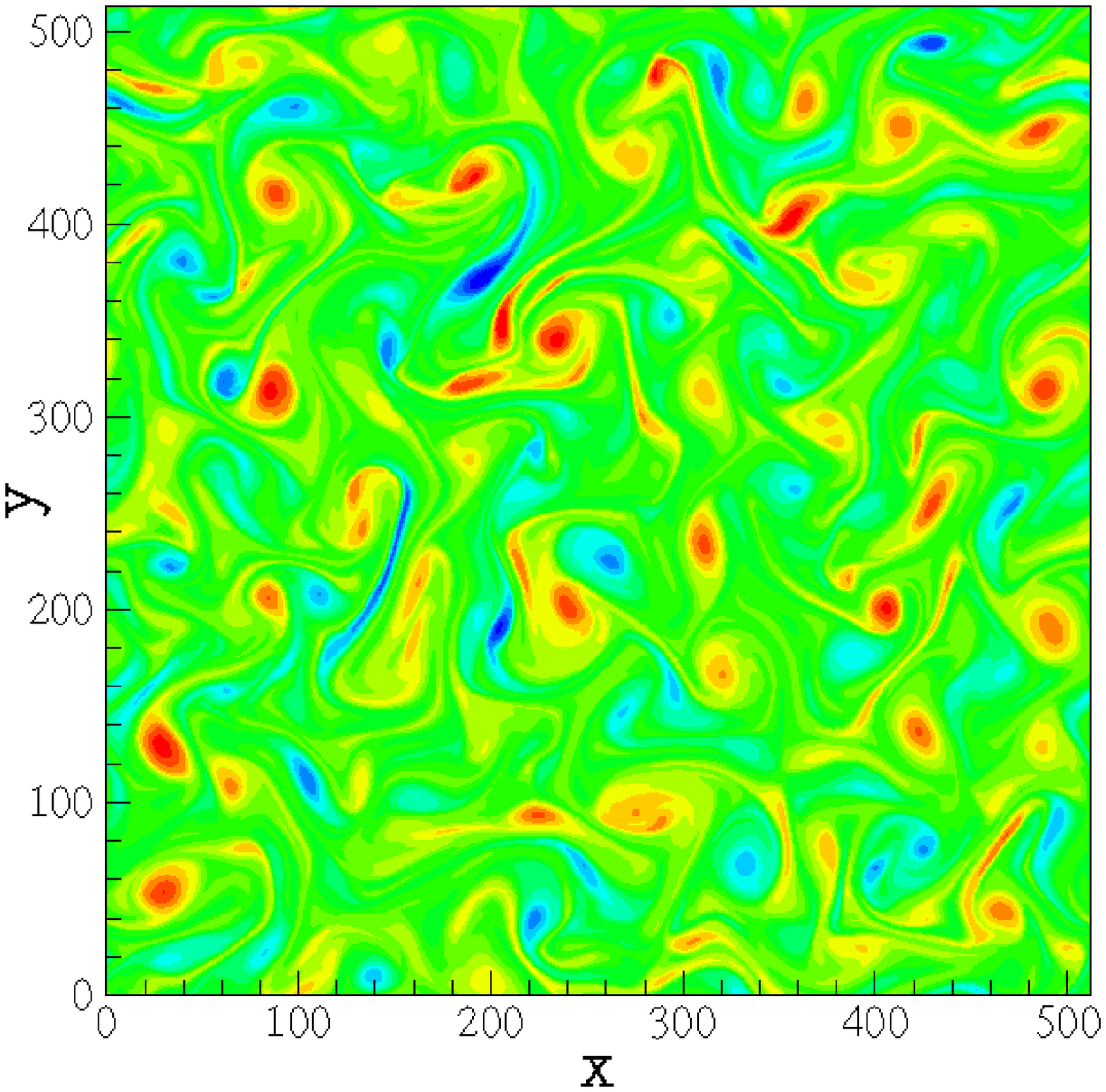} 
\end{center} 
\vspace*{-1.0cm}  
\begin{center} 
\includegraphics[scale= 0.2285833]{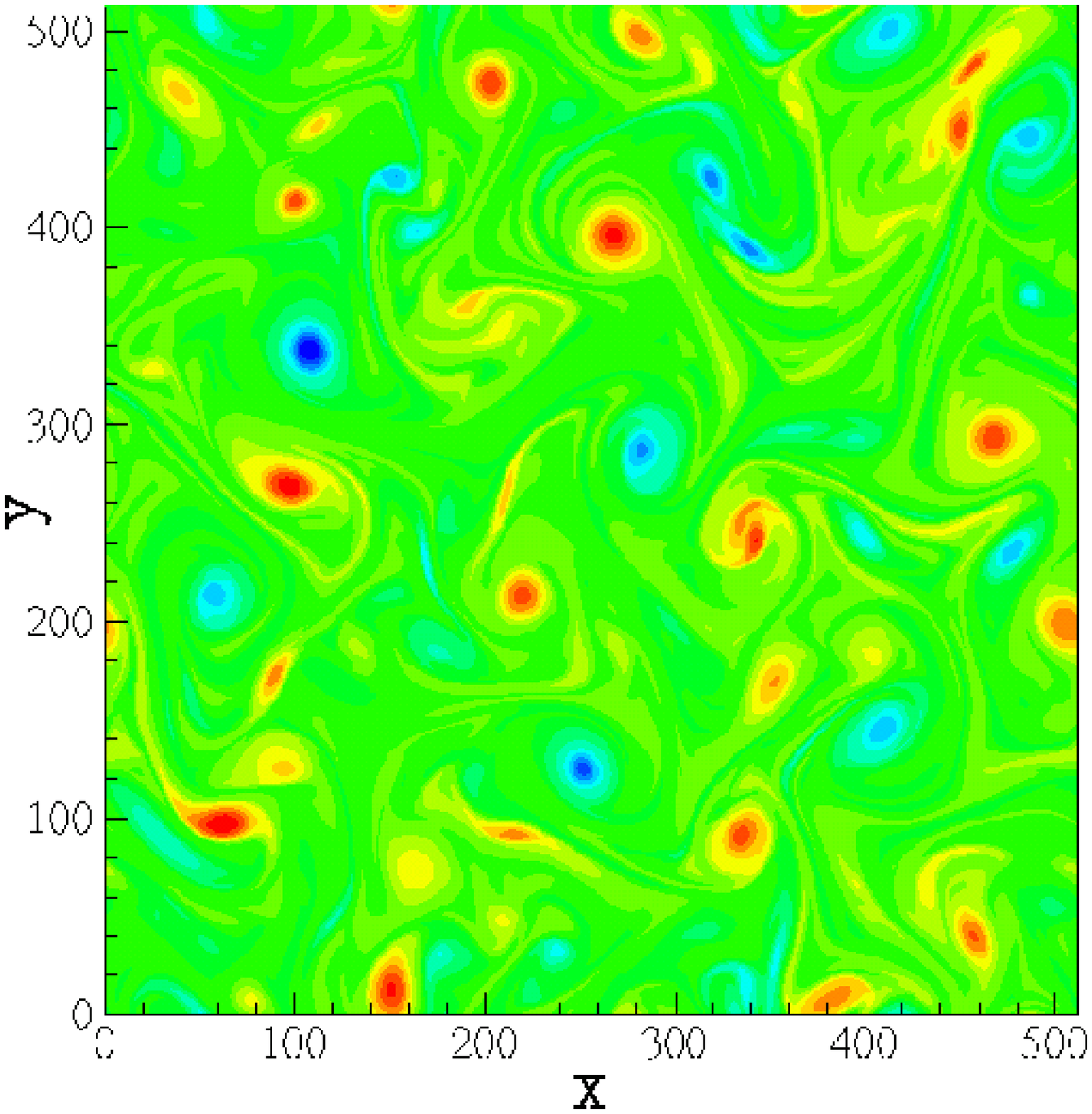} 
\hspace*{0.9cm} 
\includegraphics[scale=0.25]{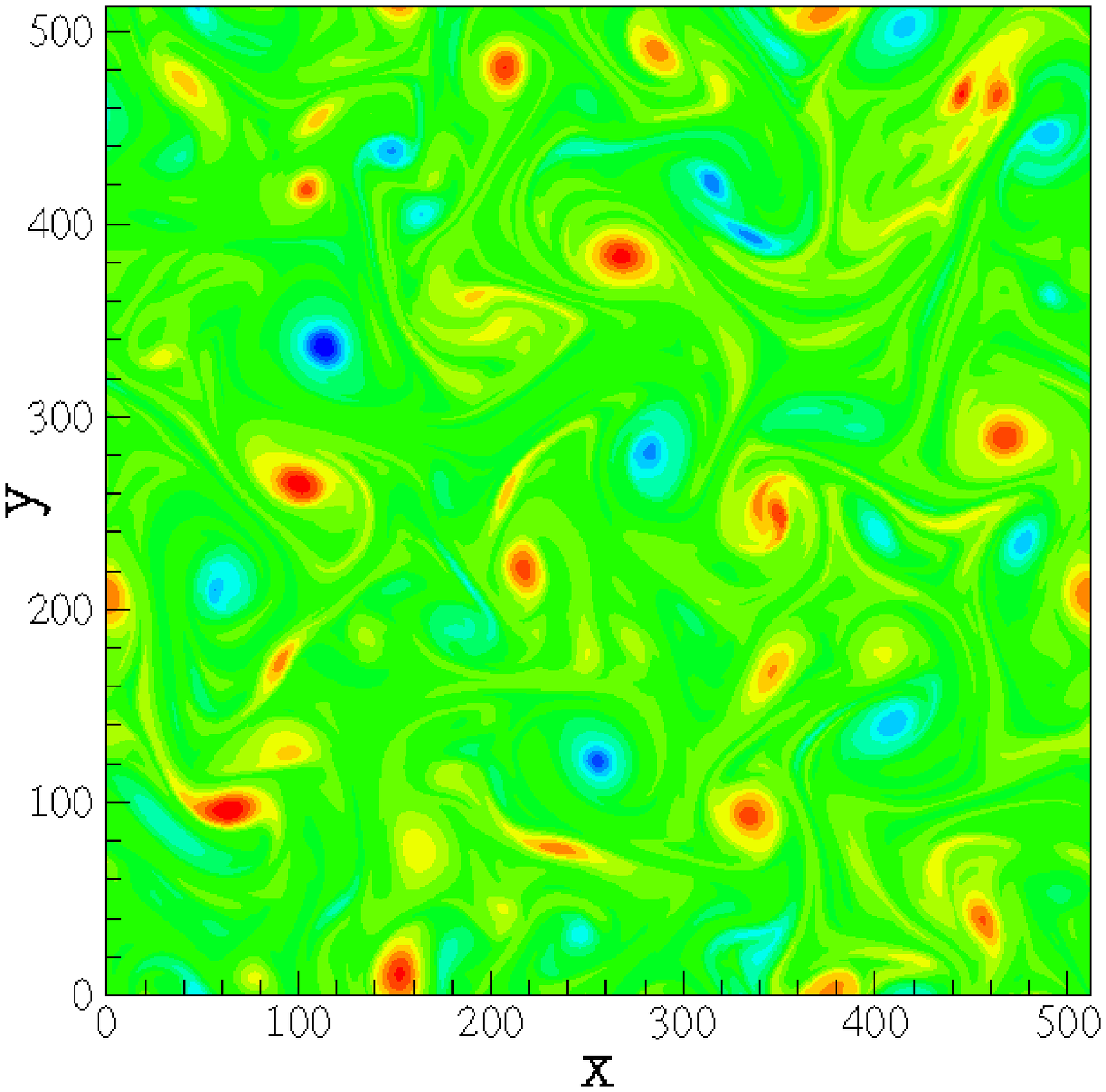} 
\end{center} 
\caption{\label{2DPlots} 
Contour plots of  vorticity at $t=0,\,5000,\,20000$:  
 ELBM results (right column), spectral results (left column). Approximate eddy turnover time is 
$t_e =700$. } 
\end{figure}

The agreement between  the two simulations is very good  up 
to  
$t/t_e \sim 20$. As, shown in  figure \ref{2DPlots}, at 
$t/t_e\sim 30$ ( i.e., $t \sim 20000$) slight difference 
between the two methods starts to 
show up in the vorticity contour plot. 
However, this difference is 
only in the small scale features of the flow, while  all 
large scales  are  still the same at this time.   
    
\begin{figure} 
\begin{center} 
\includegraphics[scale=0.225]{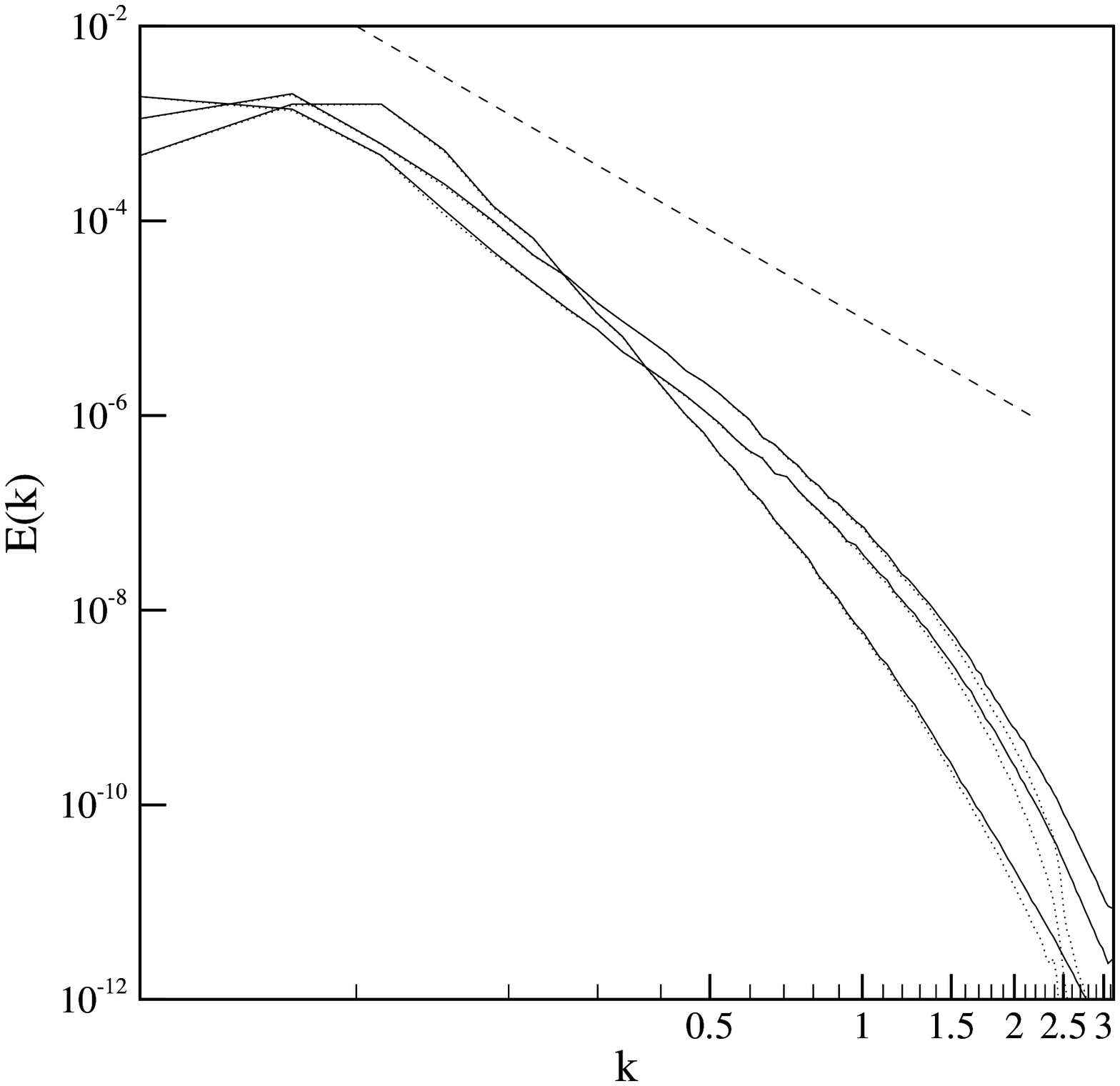} 
\includegraphics[scale=0.225]{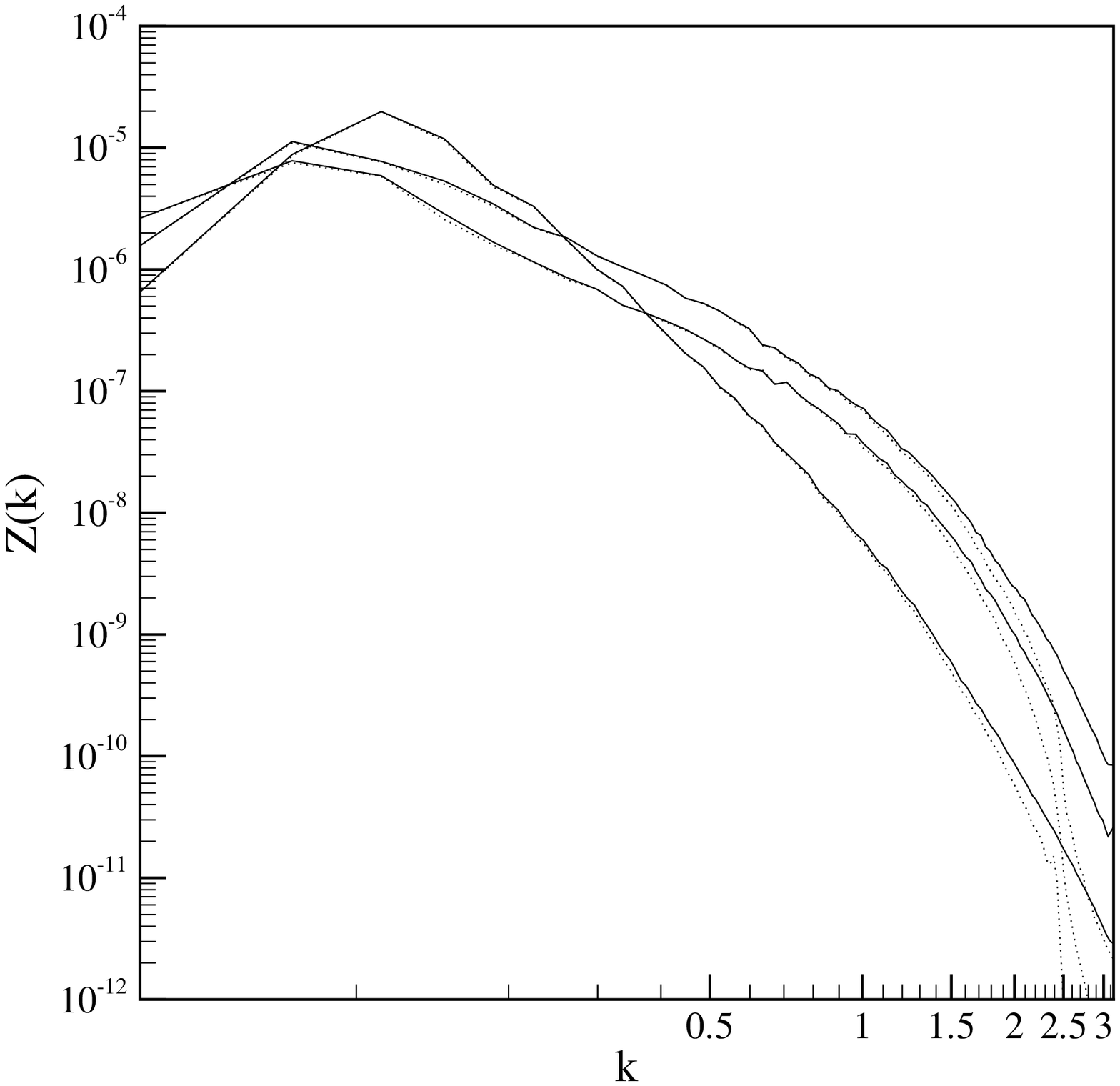}
\caption{\label{spectraE} 
Energy  and enstrophy spectra  (left and right plot respectively) in
the fully resolved simulation (example ${\rm II}$) for three 
different times: $t=1000,\,5000,\,10000$. Solid line is for  
spectral simulation and the dotted symbols represent ELB simulation.  
A dashed line showing $k^{-3}$ spectra in the energy spectra is also shown. } 
\end{center} 
\end{figure}

Figure \ref{spectraE} shows a comparison of 
the energy and enstrophy spectra, respectively, up to $t= 10000$. 
The plot shows 
that the energy and the enstrophy are narrow banded 
initially. The enstrophy and the energy plots for the SS show the 
typical behavior, where during the course of simulation high wave number 
modes start gaining enstrophy (and consequently some energy too). 

In freely decaying 2D turbulence, dissipation takes
place via a enstrophy cascade from large to small 
scales.
To prevent enstrophy pile-up, a small-scale
enstrophy sink is typically required in numerical simulations.
Figure \ref{spectraE} shows
that the ELBM naturally provides a cutoff at high wave-number.  In 
the subsequent section, we shall show that this cutoff does not affect 
the low wave-numbers region in any unphysical manner.  

\begin{figure} 
\begin{center} 
\includegraphics[scale=0.225]{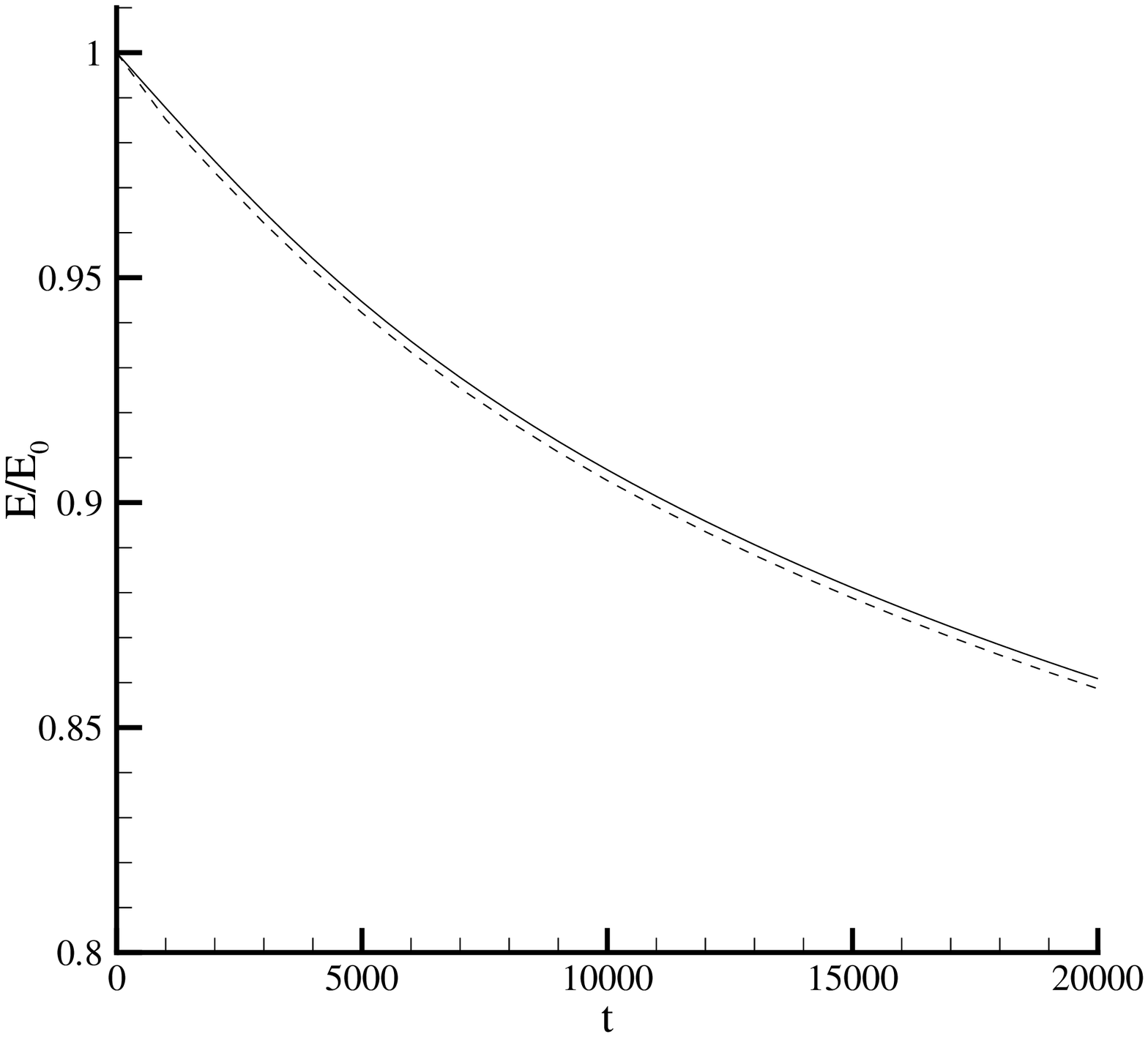} 
\includegraphics[scale=0.2250]{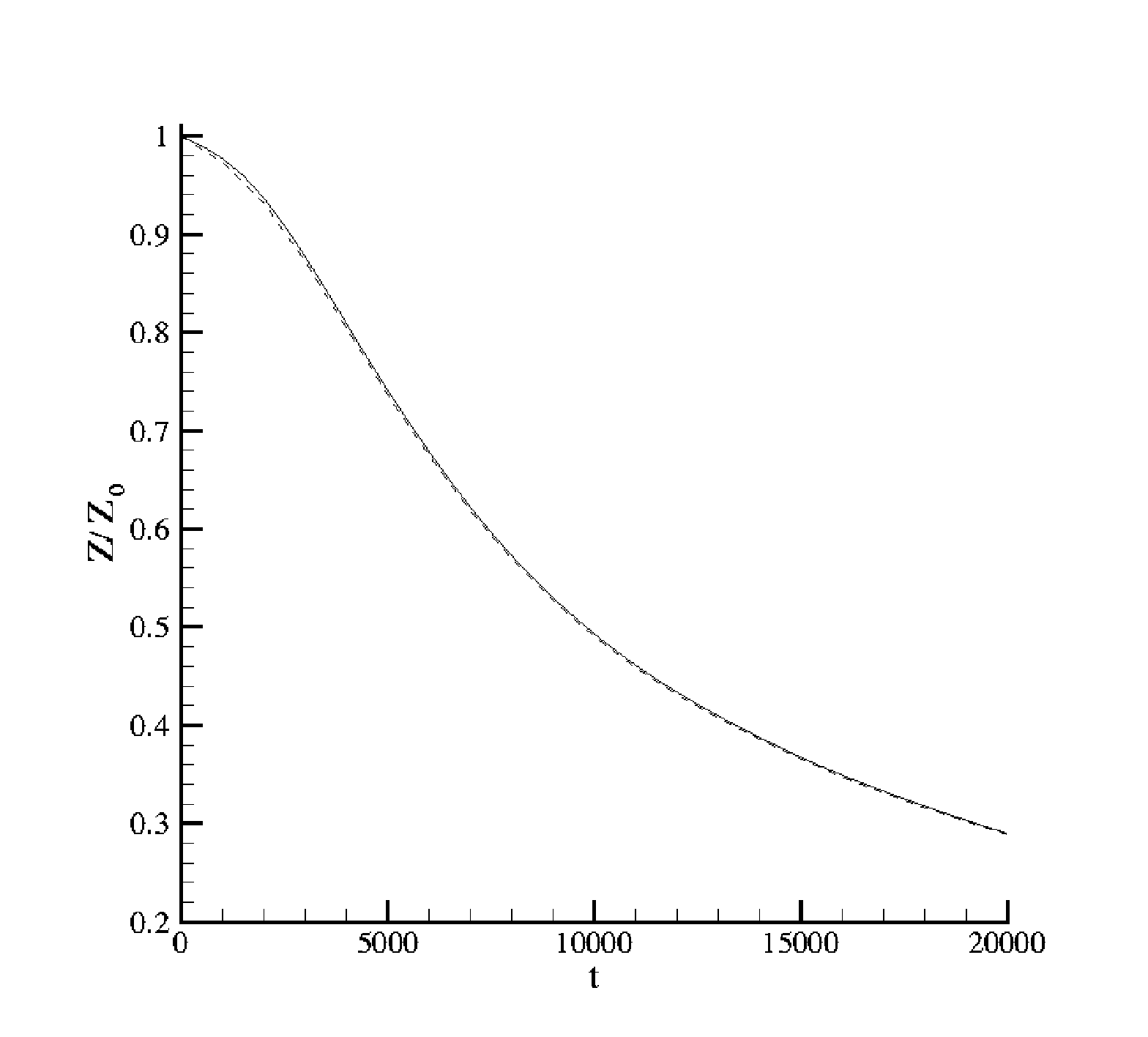} 
\caption{\label{historyE1} 
Time history of the  energy (left plot), and the enstrophy. $E_0$ and
 $Z_0$ are the energy and the enstrophy respectively at $t=0$. Solid line SS and dotted line ELBM.  
 } 
\end{center} 
% \begin{center} 
% \includegraphics[scale=0.30]{pics/historyZ1.eps} 
% \caption{\label{historyZ1} 
% Time history of the  enstrophy, solid line SS and dotted line ELBM.  
%  } 
% \end{center} 
% \begin{center} 
% \includegraphics[scale=0.30]{pics/historyP1.eps} 
% \caption{\label{historyP1} 
% Time history of the  palinsthropy, solid line SS and dotted line ELBM.  
%  } 
% \end{center} 
\end{figure}

Figure \ref{historyE1}
shows the energy and enstrophy time decay respectively.
We see that these quantities are slightly underestimated by the ELBM. 
This is probably related to the sharp cutoff at high 
wavenumber present in the ELBM simulation.  
 
As the simulation is performed in the low-Reynolds number regime, 
both methods provide a rather poor agreement with 
the Kolmogorov-Batchelor theory, and show visible
deviations from the theoretically expected spectrum $E(k)\sim k^{-3}$.  

%\FloatBarrier 
 
\subsection{Unresolved simulation: example ${\rm I}$} 

In the previous section, \ref{sect-RS}, we have verified that the
entropic method can be used to perform a fully 
resolved simulation of turbulence. 
We also observed that when the 
spectrum is not fully resolved by the computational domain, the ELBM method 
introduces a sharp dissipation at high wave numbers. 
In this section, we shall explore how 
such high-$k$ cut-off affects the spectrum of the resolved scales.  

\begin{figure} 
\begin{center} 
\includegraphics[scale=0.225]{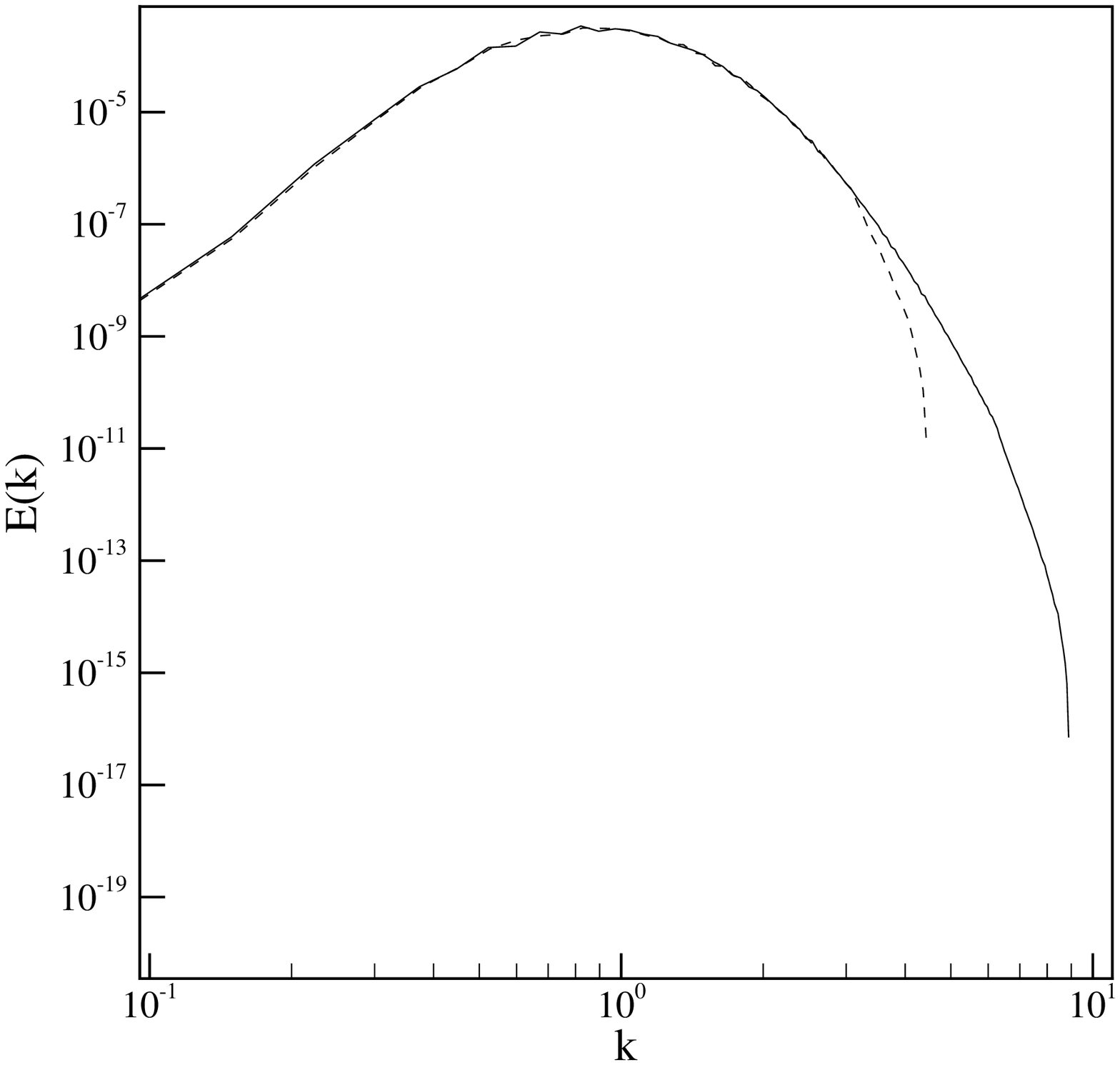} 
% \caption{\label{SpectraUI} 
% Wave number energy spectra  at $t=0$ in the  unresolved simulation 
% by ELBM.  } 
\includegraphics[scale=0.225]{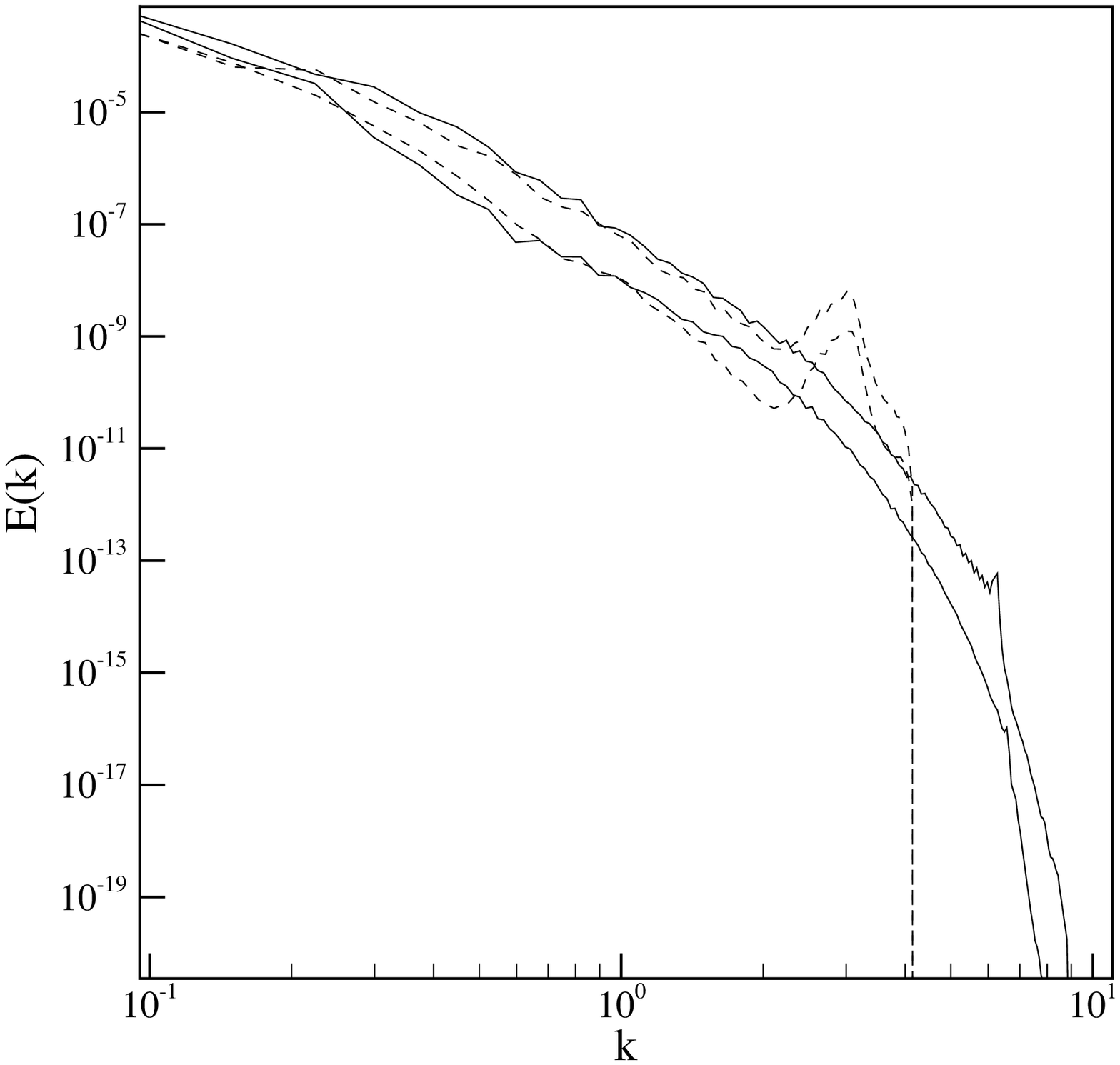} 
\caption{\label{SpectraU} 
Wave number energy spectra in the
unresolved simulation (example ${\rm I}$). The plot on 
the left hand side shows the initial condition, while that on the right 
hand side shows the energy spectra for 
three different times: $t=4000,\,20000,\,40000$. The solid line is for  
the spectral simulation and the dashed symbols represent ELB simulation } 
\end{center} 
\end{figure}

In order to magnify this effect, for the spectral simulation 
we used $256\times 256$ grid points, while for the ELB 
we performed the simulation with $128\times 128$ grid points. 
We prepare the initial condition in such a way that the 
resolved scales are the same 
for both methods (see plot on the left hand side in figure \ref{SpectraU}). 
The spectral simulation is fully resolved while the spectrum for the 
ELBM simulation is truncated almost in the middle.  
The plot of the energy spectra, 
figure \ref{SpectraU}, at 
different times shows that the ELBM simulation follows the 
spectral simulation reasonably well.  
At low wave-numbers, i.e.,  in 
the resolved scales, the deviation from the spectral 
simulation is quite small. 
Near the cutoff wavenumber, a slight 
(but localized) hump in the profile obtained from ELBM appears.  
This hump and the agreement at low wavenumbers, can be interpreted as a 
signature of the very localized nature of the ELBM dissipative
cut-off at high wavenumbers. 
 
%\FloatBarrier 
 
\subsection{Unresolved simulation: example ${\rm II}$} 
 
\begin{figure} 
\begin{center} 
\includegraphics[scale=0.225]{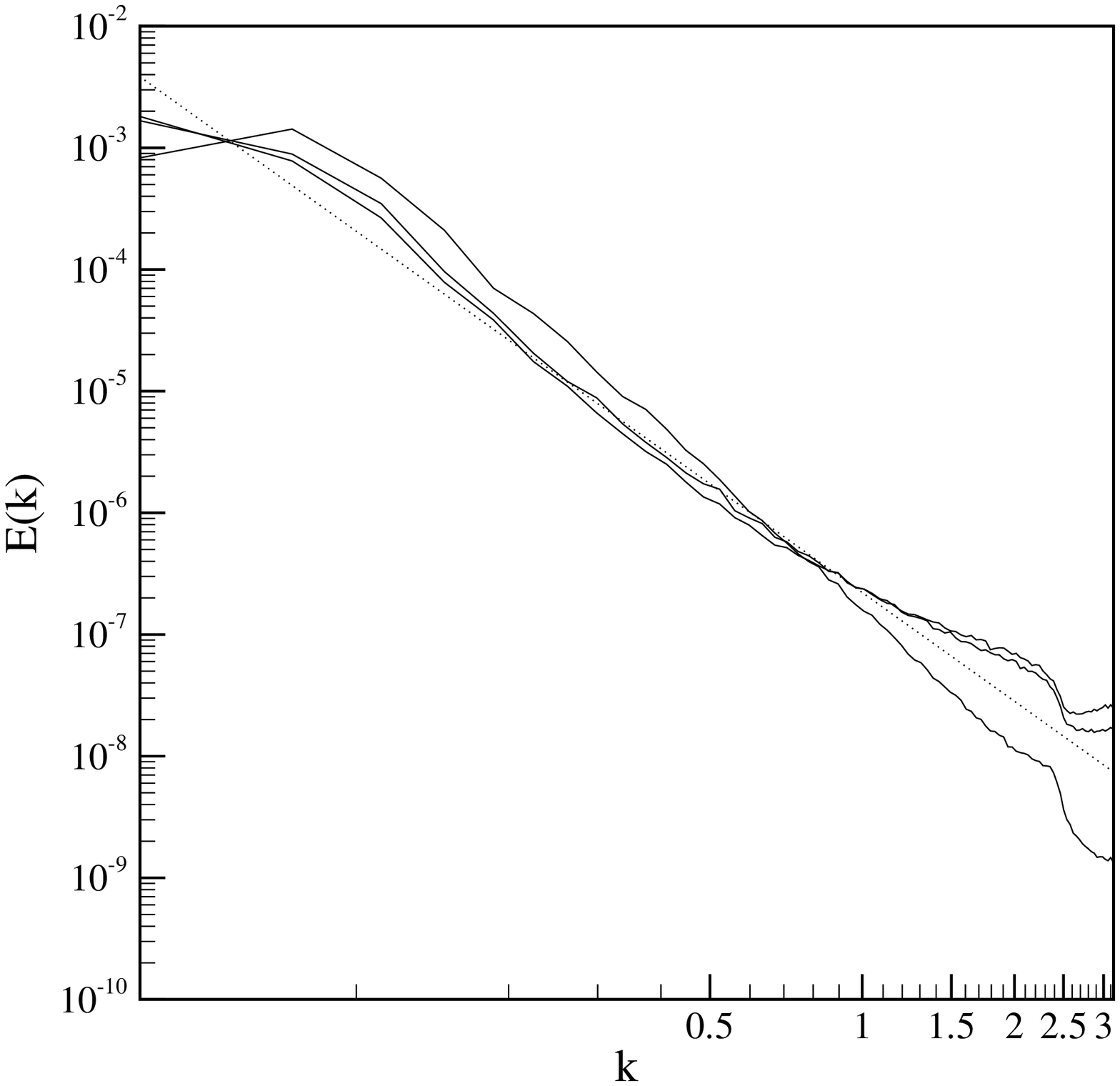} 
\includegraphics[scale=0.225]{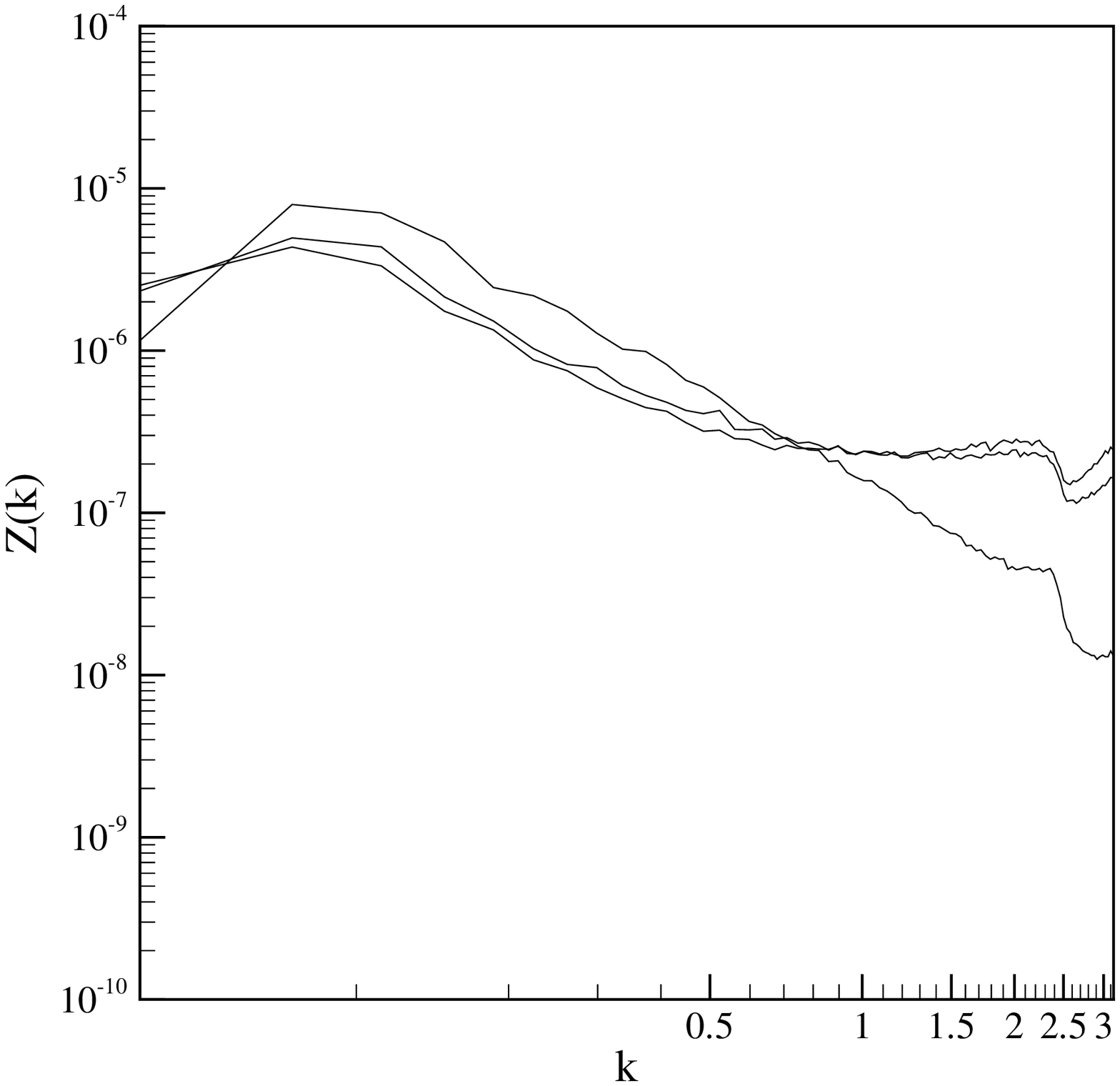}
\caption{\label{spectraE3} 
Energy  and enstrophy spectra  (left and right plot respectively) in
the  ELBM unresolved simulation (example ${\rm II}$) for three 
different times: $t=5000,\,15000,\,20000$. 
A dashed line showing $k^{-3}$ spectra in the energy spectra is also shown. } 
\end{center} 
\end{figure}

In the previous two sections, we showed that the entropic method 
can perform turbulence simulation, both resolved and underresolved, 
to a reasonable degree of accuracy. 
In this section, we shall explore the ELBM behaviour for flows 
much beyond the grid resolution.  
For this case, we have chosen the box-size as  
$L=512$ and viscosity $\nu=5.0 \times 10^{-6}$ for the ELBM 
simulation. As described earlier, the initial conditions are given in  
Fourier space with $k_0=0.2$, $A=6$ and $B=17$ (see Eq. \ref{ICE}).  
The initial velocity profile has mean kinetic energy $E(t=0)= 
1.20861512\times 10^{-4}$ and mean enstrophy as $Z(t=0)= 
1.4949915\times 10^{-6}$. A rough estimate of the eddy turnover time is $t_e 
\approx Z^{-1/2}  \approx 818$ \cite[]{Millen}. The Reynolds number 
based on the mean initial kinetic energy and the box-length as the 
characteristic length is $Re = L \sqrt{(2 E)} / \nu \approx 
1.59 \times 10^6$. The dissipation length scale is estimated as
$L_{d}\sim L \, Re^{(-1/2)} \approx 0.4$. 
Figure \ref{spectraE3} shows that during the 
time evolution, Batchelor spectra $E_k \sim k^{-3}$ for the energy and 
$k^{-1}$ for the enstrophy in the inertial regime, are reproduced to
a reasonable accuracy. 
The scaling law is best fitted at $t=15000$ as $2\times 10^{-7} k^{-3}$. 
A rough estimate of the enstrophy transfer rate yields $1.7\times 10^{-11}$.  
This gives a value of $4.8$ for the constant appearing in the 
Batchelor scaling, while the theoretical value is $2.626$. 
We remind that the determination of this constant requires an accurate
evaluation of the enstrophy transfer rate, so that we expect the agreement 
to improve with increasing  system size. 
The agreement with the scaling law in such
an underresolved simulation shows that the ELBM exhibits a sort
of built-in subgrid model for turbulent flows. 

\begin{figure}
\begin{center} 
\includegraphics[scale=0.175]{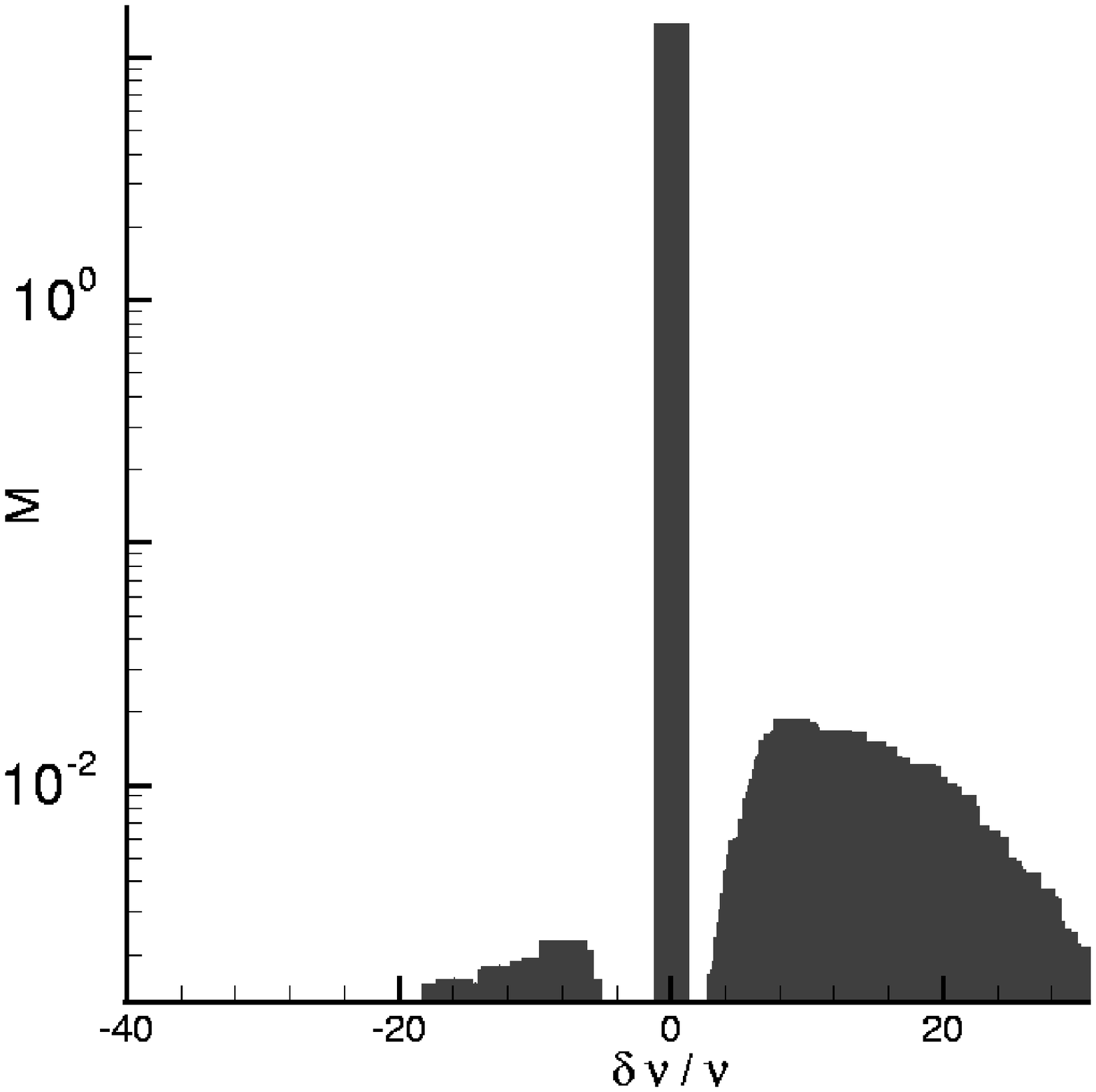}  
\includegraphics[scale=0.175]{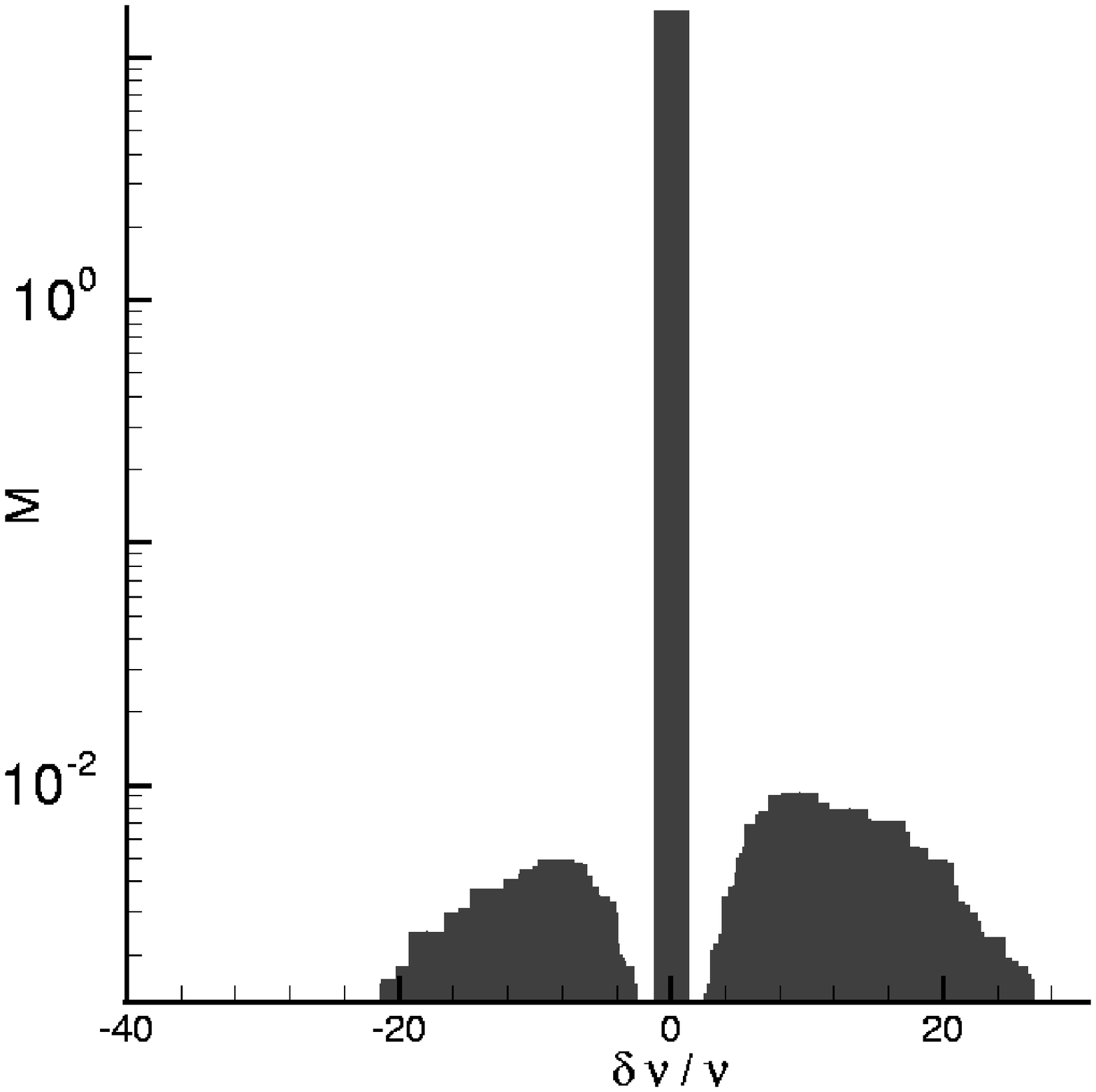} 
\includegraphics[scale=0.175]{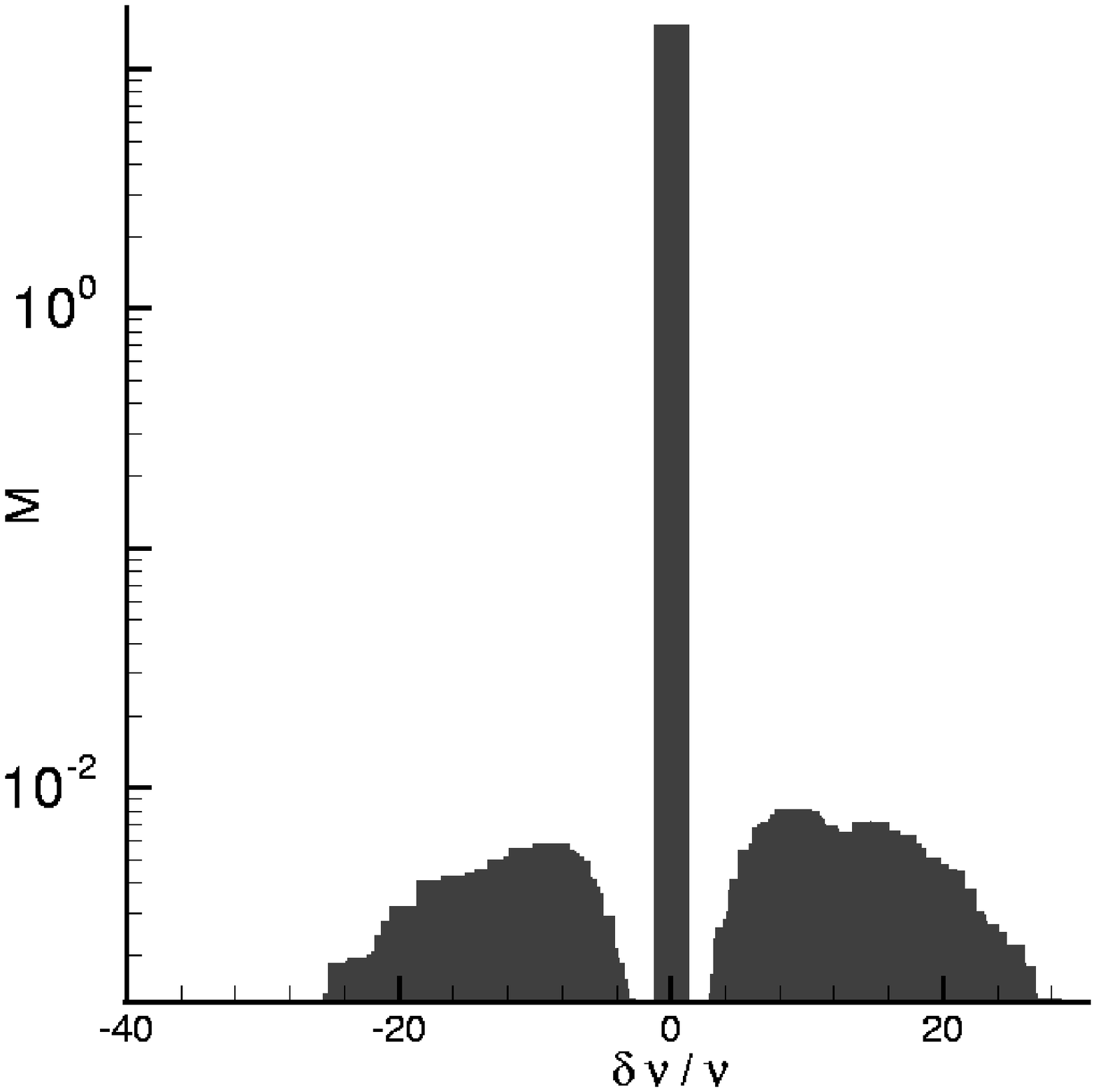} 
\end{center} 
\caption{\label{prob} 
Distribution $M$ of  $\delta \nu /\nu$ at time $t=1000, 5000, 10000$ in a
unresolved simulation. Notice the
delta function at $\delta \nu /\nu=0$, where the distribution is normalised in
such a way that area under curve is unity.
} 
\end{figure}

It is argued that in two-dimensional isotropic turbulence, the 
eddy-viscosity develops a dependence on the wavenumber and may
depart from its mean value in both positive and negative directions 
\cite[]{EddyTwoD}. It is therefore of interest to investigate whether
ELBM has similar features.
We can argue that the deviation of the effective
 relaxation frequency $\alpha \beta$ from its near equilibrium value
 $2 \beta$ is related to  the deviation of viscosity from the
 molecular viscosity $\nu(\beta) =c_s^2 ({1}/{2 \beta}-{1}/{2})$. 
In order to parameterise the relaxation behavior, we define the  `effective viscosity' of the ELBM fluid
(in lattice units) as:
\begin{equation} 
\label{VISCO} 
\nu_{\rm eff} = c_s^2 \left(\frac{1}{\alpha \beta}-\frac{1}{2}\right).
\end{equation} 
In figure \ref{prob}  the probability distribution of
 turbulent fluctuations in effective viscosity  as compared to the
bare molecular viscosity, $\delta \nu/\nu =
\nu_{\rm eff}(\alpha,\beta)/\nu(\beta)-1$ is plotted. 
From this figure a three-modal distribution is clearly visible, corresponding 
to three distinct classes of events: 

1)Plus-events, ($ \delta \nu/\nu>0$): Effective viscosity
 is much larger than  molecular viscosity.
2) Normal-events, ($\delta \nu/ \nu = 0$): Effective viscosity is identical
 to molecular viscosity, within the machine accuracy.  
3) Minus-events, ($\delta \nu/\nu <0$): Effective viscosity
 is much smaller than molecular viscosity. This case corresponds to local
micro-instability. 

The relative weights, $\left \{M_{\rm Plus} , M_{\rm Normal}, M_{\rm Minus} \right \}$, for the
three events at three different times, $t=1000,5000,10000$,  are 
$\left \{0.227002, 0.754618, 0.018380 \right \}$,
$\left \{0.099759, 0.853831, 0.046590 \right \}$, and 
$\left \{0.097823, 0.839673, 0.062504 \right \}$, respectively. This shows that, the number of Minus-events grows
in time, while Plus-events do the opposite.
The above figures have been obtained by 
defining Normal-events via the condition  $|\delta \nu/\nu|< 0.01$. 
Such a low threshold indicates that the distribution of 
Normal-events is a Dirac's delta.
This shows that the parameter $\alpha$ is highly
`intermittent', e.g. quiescent in most part of the flow, with
strong bursts of activity in small regions (this is confirmed by
visual inspection of the spatial distribution $\alpha(x,y)$, not
shown here). This type of behaviour cannot be captured
by any `conventional' picture of eddy-viscosity.

\FloatBarrier 

\section{Conclusions} 
In conclusion, we have shown that entropic lattice Boltzmann models (ELBM) 
can cope with the instabilities which plague standard lattice Boltzmann 
schemes in a regime where subgrid scales are dynamically excited. 
This boost of numerical stability is the result of securing compliance 
with the H-theorem, hence positivity of the discrete distribution function. 
Therefore it can be concluded that ELBM exhibit a built-in 
subgrid model which is rooted in a basic principle of 
statistical mechanics rather than on heuristic requirements or
high-order discretizations of the Navier-Stokes equations. 
This built-in model cannot be reduced 
to an algebraic relation between eddy-viscosity and the large 
scale shear. This is not surprising, since the present ELBM 
accounts for both positive and 
negative turbulent relaxation, the latter corresponding to local instabilities 
which escape traditional eddy-viscosity turbulence models. 
%Following the terminology of  \cite{LM}, the
%present ELBM might be classified within "pseudo-direct" simulation methods,
%namely numerical schemes which can mimic the effects
%of unresolved scales without introducing any explicit 
%turbulence modelling term in the equations.
%A typical example of "pseudo-direct" simulation is the 
%third-order upwind finite-differences scheme introduced
%by  \cite{Kawamura}.
%The distinctive mark of ELBM, as opposed to the above schemes,
%is that ELBM stems from a genuinely physical requirement and
%not by a higher-order discretization of the Navier-Stokes equations.
%As a result, and most importantly, instead of truncating the
%Navier-Stokes equations to any given order, it
%performs the resummation of an {\it infinite} number of terms in
%the lattice Knudsen number $Kn(\delta) \sim (\delta \nabla f)/f$.
Future work will tell whether such encouraging qualitative features 
can be turned into a quantitative {\it genuinely kinetic} approach 
to the problem of turbulence modeling. 
 
\bibliographystyle{natbib} 
\bibliography{jfm} 
\end{document}